\documentclass[12pt, prd, showpacs]{revtex4}
\usepackage{amssymb}
\usepackage{amsmath}

\input{tcilatex}

\begin{document}

\title{New scenarios of high-energy particle collisions near wormholes}
\author{O. B. Zaslavskii}
\affiliation{Department of Physics and Technology, Kharkov V.N. Karazin National
University, 4 Svoboda Square, Kharkov 61022, Ukraine}
\affiliation{Institute of Mathematics and Mechanics, Kazan Federal University, 18
Kremlyovskaya St., Kazan 420008, Russia}
\email{zaslav@ukr.net }

\begin{abstract}
We suggest two new scenarios of high-energy particle collisions in the
background of a wormhole. In scenario 1 the novelty consists in that the
effect does not require two particles coming from different mouths. Instead,
all such scenarios of high energy collisions develop, when an experimenter
sends particles towards a wormhole from the same side of the throat. For
static wormholes, this approach leads to indefinitely large energy in the
center of mass. For rotating wormholes, it makes possible the super-Penrose
process (unbounded energies measured at infinity). In scenario 2, one of
colliding particles oscillates near the wormhole throat from the very
beginning. In this sense, scenario 2 is intermediate between the standard
one and scenario 1 since the particle under discussion does not come from
infinity at all.
\end{abstract}

\maketitle

\section{Introduction}

During last decade, a lot of efforts was devoted to description of
high-energy collisions in the region of the strong gravitation field. This
was stimulated by the observation about possibility to obtain an
indefinitely large energy $E_{c.m.}$ in the centre of mass frame of two
colliding particles \cite{ban} (see also earlier works \cite{pir1} - \cite%
{pir3}). These findings were made for the case of rotating black holes.
Meanwhile, later, similar results were obtained for another strongly
gravitating objects. Thus, the unbounded energies $E_{c.m.}$ were found for
processes near naked singularities and wormholes. In the present article, it
is the latter case which we are interested in.

One should distinguish between two kinds of energy. The first one is $%
E_{c.m.}$ that can be measured by an observer who is present just in the
point of collision. The second one is the Killing energy $E$ measured at
infinity in the asymptotically flat space-time. In the present work, we will
discuss both of them. It is essential that if $E_{c.m.}$ is finite, $E$ is
finite as well. This was shown in \cite{inf} for the Kerr metric and in \cite%
{tangrg} for a more general case. (There is a very special case \cite{rap}
when the parameters of the metric themselves diverge, but we will not
discuss it further.) Therefore, the necessary (although not sufficient)
condition for obtaining unbounded $E$ consists in the consideration of
processes with unbounded $E_{c.m.}.$ In what follows, we will use the term
super-Penrose process (SPP), if $E$ is unbounded.

For the first time, collisions with unbounded $E_{c.m.}$ near wormholes were
considered in \cite{bambi1} for a particular type of \ wormholes, so-called
Teo wormholes \cite{teo}. They are necessarily rotating, the corresponding
space-time does not have an asymptotically flat region. In the next work, it
was shown that collisions near such wormholes can also produce unbounded $E$ 
\cite{bambi2}. Later, it was noticed \cite{kras} that high-energy collisions
can be realized even for static wormholes (for example, if two
Schwarzschild-like wormholes are glued by means of the "cut and past"
technique, see e.g. Sec. 15.2.1 of \cite{vis}). In the Krasnikov's scenario 
\cite{kras}, unbounded $E_{c.m.}$ do occur but unbounded $E$ are forbidden
since this would require the presence of the ergosphere where $E<0$.
Meanwhile, such a region is absent for the Schwarzschild-like metric. In our
previous paper it was shown that the SPP is possible for rather general
rotating wormholes \cite{wh}.

It is worth stressing that there exists nontrivial dependence between the
behavior of $E_{c.m.}(N)$, where $N$ is the lapse function, and the
existence or nonexistence of the SPP. This relation was established in \cite%
{centr}, where general classification was constructed. In doing so, $%
E_{c.m.}(N)$ itself is determined by the relative sign of radial momenta of
colliding particles. It is head-on collision that leads to the existence of
the SPP. In this sense, it is of interest to describe possible ways, how to
realize head-on collisions. \ It is this point that we make the accent on.
As far as wormholes are concerned, in previous works it was assumed that two
particles come from opposite mouths and meet near the throat (head-on
collision). Thus the corresponding experiment had "mixed" nature involving
observers from different sides of Universe.

In the present work, we suggest two completely new, alternative scenarios.
We show that an indefinitely large energies $E_{c.m.}$ and $E$ can occur
even if both particles are sent from the same side of the throat. However,
this requires two-step process. Also, we exploit the fact that in wormhole
space-times there exist bound states (impossible for black holes) when a
particle can oscillate between two turning points \cite{sd}.

One reservation is in order. All scenarios connected with using wormholes
for obtaining unbounded $E_{c.m.}$ share the same feature. Namely, the lapse
function near the throat should be small. This leads to an indefinite growth
of the curvature invariants (say, the Kretschmann scalar $K$) there.
Meanwhile, one can reconcile large $E_{c.m.}$ and $K$ remaining below the
Planckian scale by choosing the parameters of the system accordingly \cite%
{os}.

Below, we consider two types of scenarios in which high energy phenomena
reveal themselves in (i) indefinite growth of $E_{c.m.}$ (with $E$ remaining
modest), (ii) in the SPP. To this end, we consider separately (i) collision
in static spherically symmetric wormholes and (ii) in rotating axially
symmetric ones. In the first case, only $E_{c.m.}$ can be unbounded, in the
second one we explain why also $E$ can be made as large as one likes.

The paper is organized as follows. In Sec. \ref{sph} we give basic formulas
for the spherically symmetric case including the metric, equations of motion
and the energy in the center of mass of two colliding particles. In Sec. \ref%
{spscen} we describe a scenario in which one of two particles reflects from
the potential barrier, so that an ingoing particle converts into the
outgoing one. In Sec. \ref{spun} we show that, choosing in the point of
collision the metric function small enough, we can achieve indefinitely
large $E_{c.m.}$. In Sec. \ref{int} we describe another scenario in which
one of colliding particles does not come from infinity but oscillates near
the throat between turning points. In Sec. \ref{rot} we give the general
metric and equations of particle motion in the case of rotating wormholes.
In Sec. \ref{thr} we describe a general scheme of particle collisions in
such a background. In Sec. \ref{out} we analyze possible output of particles
with ultrahigh energy. In Sec. \ref{neg} we discuss the role of trajectories
with negative energy played in the high energy processes under
consideration. In particular, we discuss how they can be used in an
alternative scenario of collision. In Sec. \ref{concl} we summarize the
results and outline some perspectives.

We use the geometric system of units in which fundamental constants $G=c=1$.

\section{Spherically symmetric case: basic formulas \label{sph}}

Let us consider the spherically symmetric metric%
\begin{equation}
ds^{2}=-fdt^{2}+\frac{d\rho ^{2}}{f}+r^{2}(\rho )d\omega ^{2}\text{, }%
d\omega ^{2}=d\theta ^{2}+\sin ^{2}\theta d\phi ^{2}\text{,}
\end{equation}%
where we used a so-called quasiglobal coordinate $\rho $ (see, e.g. Sec.
3.3.2 of \cite{br}). Motion of free particles occurs in the plane which we
choose to be the equatorial one $\theta =\frac{\pi }{2}$. Equations of
motion read%
\begin{equation}
m\dot{t}=\frac{E}{f}\text{,}  \label{mt}
\end{equation}%
\begin{equation}
m\dot{\rho}=\sigma P\text{,}
\end{equation}%
\begin{equation}
m\dot{\phi}=\frac{L}{r^{2}}\text{, }  \label{phi}
\end{equation}%
where dot denotes differentiation with respect to the proper time $\tau $, $%
E $ being the conserved energy, $L\,\ $conserved angular momentum, $\sigma
=\pm 1$ depending on the direction of motion,%
\begin{equation}
P=\sqrt{E^{2}-f\tilde{m}^{2}},\text{ }  \label{zn}
\end{equation}%
\begin{equation}
\tilde{m}^{2}=m^{2}+\frac{L^{2}}{r^{2}}\text{.}  \label{mm}
\end{equation}

The forward-in-time condition $\dot{t}>0$ is satisfied, provided $E>0$.

If two particles 1 and 2 collide, one can define the energy in the center of
mass frame according to%
\begin{equation}
E_{c.m.}^{2}=-(m_{1}u_{1\mu }+m_{2}u_{2\mu })(m_{1}u_{1}^{\mu
}+m_{2}u_{2}^{\mu })=m_{1}^{2}+m_{2}^{2}+2m_{1}m_{2}\gamma \text{.}
\end{equation}%
Here, $u^{\mu }$ is the four-velocity, subscript label particles, $\gamma
=-u_{1\mu }u_{2}^{\mu }$ is the Lorentz factor of relative motion. Using (%
\ref{mt}) - (\ref{phi}) one obtains%
\begin{equation}
m_{1}m_{2}\gamma =\frac{E_{1}E_{2}-\sigma _{1}\sigma _{2}P_{1}P_{2}}{f}-%
\frac{L_{1}L_{2}}{r^{2}}\text{.}
\end{equation}%
In what follows we consider the manifold to be a wormhole. For simplicity,
we assume that the function $r(\rho )$ has one minimum at $\rho =\rho _{0}$,
so%
\begin{equation}
r\geq r_{0}\equiv r(\rho _{0})\text{.}\,
\end{equation}%
In this section, we restrict ourselves by pure radial motion $L=0$ since
this simplified case captures the main features of the phenomenon under
discussion. Then,%
\begin{equation}
\dot{\rho}=\sigma p\text{,}
\end{equation}%
\begin{equation}
p=\sqrt{\varepsilon ^{2}-f}\text{,}
\end{equation}%
where $\varepsilon =\frac{E}{m}$,%
\begin{equation}
\gamma =\frac{\varepsilon _{1}\varepsilon _{2}-\sigma _{1}\sigma
_{2}p_{1}p_{2}}{f}\text{.}  \label{ga1}
\end{equation}

\section{Scenario of collision 1: two particles come from the same mouth 
\label{spscen}}

Let us consider the following scenario. Particle 1 has the energy $E_{1}>m$
and starts its motion, say, from the right infinity. In some point it decays
to two particles 2 and 3. We assume that particle 2 has the energy $E_{2}<m$%
, whereas particle 3 has $E_{3}>m$, $\sigma _{3}=-1$. Then, particle 3
escapes to the left infinity. Meanwhile, particle 2 has the turning point $%
r_{2}$, where $p_{2}=0$, its position is given by%
\begin{equation}
f(r_{2})=\varepsilon _{2}^{2}\text{.}
\end{equation}%
We assume that $f$ is a monotonic function of $r$ in each half-space, so
there is one value of $r_{2}$ but there are two turning points in terms of $%
\rho $ in which $r(\rho )=r_{2}$. It is also clear that $f$ attains its
minimum $f_{0}$ at point $\rho _{0}$, $f_{0}=f(r(\rho _{0}))$.

Particle 2 oscillates between both turning points. Let it collide in point $%
\rho _{0}$ with one more particle 4 having (for simplicity) the same mass
that comes from infinity, $\varepsilon _{4}>1$, $\sigma _{4}=-1$. We choose
the moment of collision in such a way that particle 2 moves from the left to
the right, so $\sigma _{2}=+1$. From (\ref{ga1}), we have%
\begin{equation}
\gamma =\frac{\varepsilon _{4}\varepsilon _{2}+p_{4}(\rho _{0})p_{2}(\rho
_{0})}{f}\text{.}  \label{ga}
\end{equation}

\section{Unbounded E$_{c.m.}$ \label{spun}}

Now, we consider configurations with small $f_{0}\ll \varepsilon
_{2}<\varepsilon _{4}$. Then, $p_{2}(\rho _{0})\approx \varepsilon _{2}$, $%
p_{4}(\rho _{0})\approx \varepsilon _{4}$, 
\begin{equation}
\gamma \approx \frac{2\varepsilon _{4}\varepsilon _{2}}{f}\text{.}
\label{24}
\end{equation}%
When $f\rightarrow 0$, $\gamma $ grows unbounded, and so does $E_{c.m.}$

We would like to remind a reader that there are few scenarios of high energy
particle collisions in which unbounded $E_{c.m.}$ is obtained in head-on
collisions. The key point of such scenarios is to obtain somehow a particle
that moves in the opposite direction (with respect to another particle that
falls from infinity) and arrange collision in the point where the lapse
function is very small. This can be realized (i) near white holes \cite%
{gpwhite}, (ii) in the background of a naked singularity \cite{naked}, (iii)
in the background of a wormhole. In case (ii) there is a two-step scenario
in which a particle bounces back from an indefinitely high potential barrier
and meets a new particle coming from infinity. In case (iii), there are two
options. One of them (iii-a) consists in that two particles comes from
opposite mouths \cite{kras}. Meanwhile, in our scenario (iii-b) all
particles participating in the process, start in our universe.

Thus in our scenario we can probe the other side of a wormhole starting the
experiment on our side of it and remaining only there.

\section{Scenario of collision 2: intermediate case \label{int}}

In this section, we describe one more scenario. Let us remind a reader that
the key ingredient for obtaining unbounded $E_{c.m.}$ is a head-on collision
of two particles near the throat, under an additional condition that the
metric coefficient $f$ is small enough in the corresponding point. Thus we
have two alternatives: (i) both particles come from opposite mouths \cite%
{bambi1}, \cite{kras}, (ii) particles come from the same mouth (see above).
Meanwhile, there is also one more possibility based on the property of
wormholes having no analog in the black hole case. It was shown in \cite{sd}
that there exist states such that a particle performs bounded motion between
two turning points. Choosing an appropriate phase when particle 2 moves,
say, from the left to the right, while particle 1 comes from the right
infinity, for small $f_{0}$ we obtain the result similar to (\ref{24}) with
one difference: now $\varepsilon _{4}$ is to be replaced by $\varepsilon _{1}
$.

To make presentation self-closed, we write down the metric in the same form
as in \cite{sd}:%
\begin{equation}
ds^{2}=-dt^{2}(g(r)+\lambda ^{2})+\frac{dr^{2}}{g(r)}+r^{2}d\omega ^{2}\text{%
.}
\end{equation}

Here, for simplicity, $g=1-\frac{r_{+}}{r}$, $r\geq r_{+}$, $\lambda $ is a
constant, $r_{+}$ has the meaning of the throat radius. If $\lambda
^{2}<\varepsilon ^{2}<1+\lambda ^{2}$, a trajectory oscillates between two
turning points. Let collision occur in the phase when both particles move in
opposite directions.

Repeating our calculations step by step, we obtain for collision of
particles 1 and 2 in point $r_{0}$, moving in opposite directions radially,
the expression%
\begin{equation}
\gamma =\frac{\varepsilon _{1}\varepsilon _{2}+p_{1}(r_{0})p_{2}(r_{0})}{%
g(r_{0})+\lambda ^{2}}\text{,}
\end{equation}%
\begin{equation}
p(r)=\sqrt{\varepsilon ^{2}-(g+\lambda ^{2})}
\end{equation}%
instead of (\ref{ga}).

Choosing $r_{0}=r_{+}$, we have%
\begin{equation}
\gamma =\frac{2\varepsilon _{1}\varepsilon _{2}}{\lambda ^{2}}\text{.}
\end{equation}

If $\lambda $ is sufficiently small, $\gamma $ can be made as big as one
likes.

Such a scenario can be thought of as an intermediate case between the
aforementioned scenarios in the sense that particle 2 comes neither from the
left infinity nor from the right one. It was present near the throat because
of initial conditions. And, this scenario 2 has advantage as compared to
scenario 1 in that we should not arrange two-step process. It is sufficient
to arrange one-step collision.

\section{Rotating wormholes \label{rot}}

Now, we consider a more general metric that takes into account the effect of
rotation:

\begin{equation}
ds^{2}=-N^{2}dt^{2}+g_{\phi }(d\phi -\omega dt)^{2}+\frac{d\rho ^{2}}{A}%
+g_{\phi }d\theta ^{2}\text{,}
\end{equation}%
where the coefficients do not depend on $t$ and $\phi $, $\omega >0$. (To
simplify formulas, we use notation $g_{\phi }$ for the component of the
metric tensor $g_{\phi \phi }$). We suppose that the equatorial plane is a
plane of symmetry and are interested in the motion within this plane only.
Instead of (\ref{mt}) - (\ref{phi}), equations of motion read now%
\begin{equation}
m\dot{t}=\frac{X}{N^{2}}\text{,}
\end{equation}%
\begin{equation}
m\frac{N}{\sqrt{A}}\dot{\rho}=P_{r}=\sigma P\text{,}
\end{equation}%
\begin{equation}
m\dot{\phi}=\frac{L}{g_{\phi }}+\frac{\omega X}{N^{2}}\text{, }
\end{equation}%
where 
\begin{equation}
X=E-\omega L\text{,}  \label{x}
\end{equation}%
\begin{equation}
P=\sqrt{X^{2}-N^{2}\tilde{m}^{2}},\text{ }
\end{equation}%
\begin{equation}
\tilde{m}^{2}=m^{2}+\frac{L^{2}}{g_{\phi }}\text{.}
\end{equation}

The forward-in-time condition gives us%
\begin{equation}
X\geq 0\text{.}  \label{ft}
\end{equation}

We assume that our metric has a wormhole character. This means that $g_{\phi
}$ has a minimum in some point $\rho _{0}.$ For simplicity we assume that $N$
has also minimum in this point, $N(\rho _{0})\neq 0$ and $N(\rho _{0})\ll 1$.

\section{Collisions near throat of rotating wormhole: scenario 1 \label{thr}}

Again, we consider the two-step scenario. Our aim is to elucidate, whether
or not the energy extraction from a wormhole is possible and whether or not
it can be unbounded. In general, energy gain in this context is nothing else
than the Penrose process \cite{pen}. Let us repeat that, if it is formally
(in the test particle approximation) unbounded, it is called the
super-Penrose process (SPP). If the Penrose process is realized in the
scenario that involves collision, it is called the collisional Penrose
process (for black holes, this process is reviewed in \cite{jer}). On the
first stage, particle 1 decays to particles 2 and 3. Particle 3 escapes to
the left infinity while particle 2 moves to the right. Both $E_{2}>0$ and $%
E_{3}>0$. On the second stage, particle 4 comes from infinity and collides
with particle 2 near the throat. This is head-collision like in the static
case. As a result, particles 5 and 6 are created. We assume that $E_{5}<0$
and $E_{6}>0$, particle 6 escapes to the right infinity.

Here, there are two essential differences now as compared to the static
case. First, we assumed that the ergoregion does exist that makes it
possible to have $E<0$. Such option was forbidden in the limit $\omega
\rightarrow 0$ corresponding to the static metric. Second, we cannot put all
angular momenta equal to zero. Moreover, some of them should be large.

To explain this, let us consider the conservation laws for the energy and
angular momenta. 
\begin{equation}
E_{2}+E_{4}=E_{5}+E_{6}\text{,}  \label{e}
\end{equation}%
\begin{equation}
L_{2}+L_{4}=L_{5}+L_{6}\text{.}  \label{L}
\end{equation}

It follows from (\ref{e}) and (\ref{L}) that%
\begin{equation}
X_{tot}\equiv X_{2}+X_{4}=X_{5}+X_{6}\text{.}  \label{x56}
\end{equation}%
As, by assumption, $E_{5}<0$ and $E_{2}>0$, $E_{6}>E_{4}$. Eq. (\ref{ft})
with $\omega >0$ entails that $L_{5}<0$.

Further, we want to have $E_{6}$ large positive, so $E_{5}$ should be large
negative. Formally, $E_{5}\rightarrow -\infty $, $E_{6}\rightarrow +\infty $%
. Meanwhile, as all energies and angular momenta of particles on the 1st
stage are supposed to be finite, the quantities $X_{2}$ and $X_{4}$ are
finite as well. Taking into account that $X_{5}>0$ and $X_{6}>0$ for the
same reason (\ref{ft}), we see that each of them should be finite according
to (\ref{x56}). Therefore, we want to have configurations with $%
L_{5}\rightarrow -\infty $, $L_{6}\rightarrow +\infty $. Thus divergences in
the right hand sides of (\ref{e}) and (\ref{L}) should compensate each
other. It is seen from (\ref{x}) that $E_{6}=X_{6}(\rho _{0})+\omega (\rho
_{0})L_{6}$. Then, for finite $X_{6}(\rho _{0}),\omega (\rho _{0})$ and $%
L_{6}\rightarrow +\infty $, the energy $E_{6}\rightarrow +\infty $ as well.
This realizes the super Penrose process, when the energy $E$ detected by an
observer at infinity is as large as one likes.

\section{Output of collision \label{out}}

In Section \ref{thr}, we outlined the desired features of the process, but
the question remained, whether or not it can be realized. In principle,
further analysis is required that, apart from the conservation of the energy
and angular momentum, takes into account also the conservation of the radial
momentum. This is the most subtle and crucial point. Happily, there is no
need in carrying out such analysis here since we reduced the problem to the
one that has been already \ investigated in \cite{wh} and generalized in 
\cite{centr}. Namely, the following statement was proved there.

Let (i) two particles collide in the point where $N\ll 1$ but the horizon is
absent (as it takes place for the wormhole metric). Then, (ii) for
head-collision the energy of an escaping particle is not bounded. But both
these conditions are fulfilled now in our scenario. The aim of the 1st stage
consisted in the possibility to prepare particle 2 that moves from the left
to the right. On the 2nd stage high-energy head-on collision does occur.

It is worth stressing that both for a wormhole and a naked singularity the
dependence $E_{c.m.}(N)$ for small $N$ has the same form $E_{c.m.}(N)\sim
N^{-1}$ and this gives rise to unbounded $E$ - see line 3 in Table 1 on page
6 in \cite{centr}. Independently of origination of head-on collision near
the throat with very small $N$, once it occurred, it leads to unbounded
energies at infinity $E$.

\section{Trajectories with negative energies and scenario 2 \label{neg}}

The key role in the scenario under discussion, as well as in any Penrose
process, is played by the states with negative energy. Strange as it may
seem, only quite recently the properties of such trajectories were
elucidated and described in \cite{gpthr} for the Kerr metric. Later on, they
were generalized in \cite{neg}. It turned out that corresponding geodesics
cannot stay forever in the region external with respect to the horizon. The
complete curve inevitably crosses the horizon. Correspondingly, a particle
with $E<0$ cannot oscillate between two turning points outside the horizon
or move on the circular orbit. (The similar statements are valid for the
Reissner-Nordstr\"{o}m black hole \cite{ch}.) In the wormhole case there is
no horizon and the situation changes drastically. The particle with $E<0$
cannot escape to either of two infinities. Therefore, it must oscillate
between turning points.

Thus in our scenario, after the 1st collision, one of particles sits on the
trajectory with $E<0$ and in the phase when it moves outward, it collides
with a particle coming from infinity, creating a new particle with
indefinitely large energy.

From another hand, trajectories of such a type can be used for one more
scenario of collision. Omitting formulas, we describe it qualitatively. If a
particle has energy $E<0$, it cannot come from infinity or escape to
infinity. Instead, it oscillates between turning points. Let collision
between particle 1 coming from infinity and particle 2 oscillating inside a
wormhole occur when they move in opposite directions. The reaction can be
described as $1+2\rightarrow 3+4.$ If the lapse function in the point of
collision (say, exactly in the throat) is small enough, we again obtain
indefinitely large $E_{3}$, provided $L_{3}$ is big and negative. The
essential difference between this scenario and the one described above in
Sec. \ref{thr} consists in that there is no need in a two-step process.

\section{Conclusions \label{concl}}

One of the methods of obtaining the super-Penrose process consists in
arranging the head-on collision in the point with a small value of the lapse
function. To this end, a particle that was ingoing converts into an outgoing
due to reflection from the potential barrier with subsequent collision with
another particle coming from infinity. This is realized in the metric with
naked singularities \cite{inf}, \cite{tangrg}, \cite{naked} where the
potential barrier has indefinitely big height. Meanwhile, in the present
work we considered wormholes, the potential barrier being finite.

In the present work, we suggested two new scenarios. In scenario 1, both
particles are sent from infinity from the same side of a wormhole. It turned
out that two main features are inherent to this scenario. For pure static
wormholes, it warrants unbounded $E_{c.m.}$ If a wormhole is rotating, it
also leads to unbounded $E$, i.e. the super-Penrose process. A separate
question arises, how a remote observer who registers high-energy particles
at infinity, can distinguish between a naked singularity and a wormhole.

In scenario 2, particle 1 comes from infinity while particle 2 oscillates
between turning points from the very beginning. It can be considered as an
intermediate scenario between a standard one (when both particles come from
different mouths) and scenario 1 outlined above. In doing so, particle 2
does not come from infinity at all.

It turns out that trajectories with finite motion near the wormhole throat
can play a double role. First, they can serve as initial conditions in
collisions leading to unbounded $E_{c.m.}$ Second, after collisions, one of
product of reaction can sit on such a trajectory. Thus, either motion along
a trajectory under discussion can be specified as some initial condition or
a particle can appear there as a result of a previous collision. Anyway, one
cannot determine the origin of such a trajectory without additional
assumptions.

All discussion was carried out in the test particle approximation. As long
as the energy value does not exceed the parameters of the metric, this looks
quite reasonable. Say, in the case of Kerr-like wormholes with the
parameters $M$ and $a$, one can obtain the energy $1\ll \frac{E}{m}\ll a,$ $%
M.$ Especially interesting is to make attempt of finding self-consistent
solutions with the backreaction taken into account but this problem is
beyond of our task.

\begin{acknowledgments}
The work is performed according to the Russian Government Program of
Competitive Growth of Kazan Federal University.
\end{acknowledgments}

\end{document}